\documentclass[twocolumn,showpacs,preprintnumbers,aps]{revtex4}

\newcommand{\BE}{\begin{equation}}
\newcommand{\EE}{\end{equation}}
\newcommand{\BA}{\begin{eqnarray}}
\newcommand{\EA}{\end{eqnarray}}

\newcommand{\BW}{\begin{widetext}}
\newcommand{\EW}{\end{widetext}}

\input epsf.tex

\begin{document}


\title{Sound generated by rubbing objects}
\author{Zhen Ye}\affiliation{Wave Phenomena Lab, Department of Physics, National
Central University, Chungli, Taiwan, R. O. China}
\date{\today}

\begin{abstract}

In the present paper, we investigate the properties of the sound
generated by rubbing two objects. It is clear that the sound is
generated because of the rubbing between the contacting rough
surfaces of the objects. A model is presented to account for the
role played by the surface roughness. The results indicate that
tonal features of the sound can be generated due to the finiteness
of the rubbing surfaces. In addition, the analysis shows that with
increasing rubbing speed, more and more high frequency tones can
be excited and the frequency band gets broader and broader, a
feature which appears to agree with our intuition or experience.

\end{abstract}

\pacs{ 43.30.Ma, 43.30.Nb}

\maketitle

Rubbing objects is an everyday experience. People who have gone
through cold winters may have rubbed hands to get warm. Rubbing
also occurs in Nature. The earthquake rupture is one of the most
familiar but disastrous ones. A common observation is that rubbing
can generate sound or noise. The sound generated by rubbing hands
is certainly common to nearly every one. The sound generation by
ice floe rubbing in the Arctic ocean contributes significantly to
the ocean ambient noise\cite{PT,Xie}, and may also help sea
animals in finding appropriate holes or breaking segments in ice
to breath. The sound generated by rubbing scoops against frying
pans, however, is likely something most people would like to
avoid.

In the present paper, we wish to consider the sound generated by
rubbing two finite objects. A model is proposed for the sound
generation due to the roughness of the contacting surfaces of the
two objects. The sound field is calculated and is analyzed for its
relation to the roughness and the rubbing speed. It is shown that
sound field can be expressed in terms of a series of eigen-mode
excitations, leading to tonal features. It is suggested that when
the rubbing speed is low, only low frequency modes are possibly
excited. With increasing speed, more and more high frequency modes
can be excited, making the frequency spectrum broader. These
features appear to be in accordance with our intuition or daily
experience.

\begin{figure}[hbt]
\vspace{10pt} \epsfxsize=1.5in\epsffile{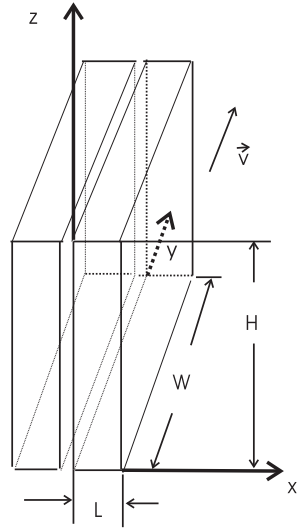}\smallskip
\caption{Conceptual layout of the system.} \label{fig1}
\end{figure}

Consider the problem of two object rubbing. The conceptual layout
is shown in Fig.~\ref{fig1}. One of the surfaces is moving with a
constant velocity along the positive $y$ axis, while the other is
assumed to be at rest. For simplicity, we assume that the two
objects are identical; for different sizes, the effective rubbing
surface will equal that of the smaller one. The shape of the
objects is a cubic with thickness $L$, width $W$ and height $H$
along the $x$-, $y$-, and $z$-th axes respectively.

When rubbing occurs, a shear displacement field will be generated
at the surfaces and is revealed as the shear waves. Such shear
waves can leak out through defects or radiate out at the
boundaries, and transform into the sound we hear or record by
machines, e.~g. microphones.

The governing equation for the shear waves inside the objects is
\BE \left(\frac{\partial^2}{\partial t^2} - c_s^2 \nabla^2\right)
u(x,y,z,t) = 0, \label{eq:we}\EE where $c_s$ is the shear speed of
the objects.

The boundary conditions are \cite{Bre} \BE \left. \frac{\partial
u}{\partial z}\right|_{z=0, H} =0, \ \ \left. u\right|_{y=0,W} = 0
\label{eq:bc1},\EE and \BE \left. \frac{\partial u}{\partial
x}\right|_{x=L} = 0, \ \  \left. \mu\frac{\partial u}{\partial
x}\right|_{x=0} = F(y,z,t), \label{eq:bc3} \EE where $\mu$ is the
shear modulus of the material.

The stress can be separated into two parts: the mean and the
variation, {\it i.~e.} $ F(t, y,z) = \langle F(t,y,z)\rangle +
\Delta F(y,z,t)$; here $\langle \cdot\rangle$ denotes the ensemble
average. We assume that $\langle F(t,y,z)\rangle$ is independent
of $y,z$. It can be shown that the averaged stress can only excite
the uninteresting zero-mode, and will thus be ignored.

By Fourier transformation, \BE u(x,y,z,t) = \int d\omega
e^{-i\omega t} \hat{u}(x,y,z,\omega). \label{eq:ft}\EE
Eq.~(\ref{eq:we}) becomes \BE (\nabla^2 + k^2_s) \hat{u}(x,y, z,
\omega) = 0, \label{eq:we1}\EE with $$k_s = \frac{\omega}{c_s}.$$
The general solution to Eq.~(\ref{eq:we1}) with
Eqs.~(\ref{eq:bc1}) and (\ref{eq:bc3}) is \BW\BE
\hat{u}(x,y,z,\omega) = \sum_{n=0,m=0}^{\infty}
A_{mn}\cos(k_{mn}(x-L))\sin(k_my)\cos(k_nz), \label{eq:s1}\EE\EW
where
$$k_n = \frac{n\pi}{H},\ \ k_m = \frac{m\pi}{W}, \ \ k_{mn} = \sqrt{k_s^2 - k_m^2 -
k_n^2},$$ and $m,n$ are positive integers.

The coefficient $A_{mn}$ is determined from \BE \sum_{m,n}
k_{mn}A_{m,n}\sin(k_{mn}L)\sin(k_my)\cos(k_nz) = \frac{1}{\mu}
F(\omega,y,z).\EE Then $A_{mn}$ is solved as
\begin{widetext}\BE A_{mn} = \frac{4}{HW[1+\delta_{n,0}][1+
\delta_{m,0}]}\frac{1}{k_{mn}\sin(k_{mn}L)} \frac{1}{\mu} \int_0^H
dz\int_0^W dy F(\omega,y,z)\sin(k_my)\cos(k_nz), \label{eq:s2}\EE
\end{widetext} with $F(\omega,y,z) = \frac{1}{2\pi}\int dt e^{i\omega t}F(t,
y,z).$ It is easy to verify that  $\langle A_{mn}\rangle$, caused
by the averaged stress, is proportional to
$\delta_{n,0}\delta_{m,0}.$ Therefore only zero modes are possible
for constant stresses. We will not discuss this situation.

The purpose to calculate the intensity of the shear field which is
related to the relevant sound field. That is, we need calculate
the correlation function \BE D(x,y,z,t; x',y',z',t') \ \equiv\
\langle u(x,y,z,t)u^\star(x',y',z',t')\rangle.  \label{eq:uc}\EE
From the above derivation, it is clear that the key is to find the
correlation function of the fluctuation part of the stress at the
surface, $\langle \Delta F(y,z,t) \Delta F(y',z',t')\rangle $.
Obviously the fluctuation is caused by the roughness of the
contacting surfaces. If we assume that the roughness is
homogeneous and completely random, i.~e. spatially uncorrelated at
different point when the system is at rest. When the surface is
moving against each other along the $y$-axis, the spatial
separation along this direction is correlated at a later time
determined as $\Delta/v$. This consideration leads to \BE
\langle\Delta F(y,z,t) \Delta F(y',z',t')\rangle = S \delta(z-z')
\delta(y-y' - v(t-t')), \label{eq:fc} \EE where $S$ is a strength
factor.

Applying Eq.~(\ref{eq:fc}), the intensity field can be calculated.
The procedure is to substitute Eqs.~(\ref{eq:s1}) and
(\ref{eq:s2}) into Eq.~(\ref{eq:ft}), then calculate
Eq.~(\ref{eq:uc}) by taking into account of Eq.~(\ref{eq:fc}). We
finally get \BW\BA <u(x,y,z,t)u^\star(x,y,z,t)> &=& S\int d\omega
\sum_n \cos^2(k_nz) \sum_{m,m'}
\cos(k_{mn}(x-L))\sin(k_my)\cos(k_{m'n}(x-L))\sin(k_{m'}y) \times
\nonumber\\
& &     \frac{H}{2}C_{mn,m'n} \frac{k_m}{(\frac{\omega}{v})^2 -
k_m^2}\left((-1)^m e^{i\frac{\omega}{v}W} -
1\right)\frac{k_{m'}}{(\frac{\omega}{v})^2 -
k_{m'}^2}\left((-1)^{m'} e^{i\frac{\omega}{v}W} - 1\right).\EA\EW
This equation can be rewritten as \BW\BE
<u(x,y,z,t)u^\star(x,y,z,t)> = \frac{S}{\mu^2}\int d\omega\sum_{n}
\cos^2(k_nz)|Q_{n}(x,y,\omega)|^2,\EE where \BE Q_n(x,y,\omega) =
\sum_{m} \frac{2}{W\sin(k_{mn}L)} \cos(k_{mn}(x-L))\sin(k_my)
\frac{k_m}{(\frac{\omega}{v})^2 - k_m^2}\left((-1)^m
e^{i\frac{\omega}{v}W} - 1\right).\label{eq:Q}\EE\EW Therefore the
frequency spectrum of the sound intensity field generated by
rubbing is \BE {\cal P}(\omega) = \frac{S}{\mu^2}\sum_{n}
\cos^2(k_nz)|Q_{n}(x,y,\omega)|^2,\EE with $Q_n$ being given by
Eq.~(\ref{eq:Q}).

A few general features can be observed from the spectral formula.
First, the strength factor $S$ controls the overall sound
intensity level. It is expected to depend on a few parameters of
the surfaces, such as the friction coefficients and the mechanical
properties of the surfaces. Second, due to the factor
$\sin(k_{mn}L)$ in the denominator of $Q_n$, the resonance feature
is expected to appear when $\sin(k_{mn}L) = 0$, leading to the
phenomenon of tonal sounds. This also shows that the thickness
mainly defines the resonance feature. Next, the rubbing speed $v$
enters $Q_n$ in the form of $\omega/v$ in the denominator. For a
fixed frequency, the decreasing of moving speed will reduced the
strength of $Q_n$. Therefore with decreasing speed, high frequency
components tend to decay accordingly. Furthermore, since $k_{mn} =
\sqrt{(\omega/c_s)^2 - (m\pi/W)^2 - (n\pi/H)^2}$, the cut-off
frequencies are determined by $(\omega/c_s)^2 - (m\pi/W)^2 -
(n\pi/H)^2 \geq 0.$ These features comply with the intuition or
experience. It is apparent that the present model can be extended
to consider other rubbing surfaces, by adjusting the correlation
function of the surface stress. All these features deduced from
the theory tend to support the experimental observation of the
sound generated by ice-floe rubbing\cite{Xie}. We note that the
experimental results have been interpreted by an alternative model
\cite{Ye}.

\begin{figure}[hbt]
\vspace{10pt} \epsfxsize=2.5in\epsffile{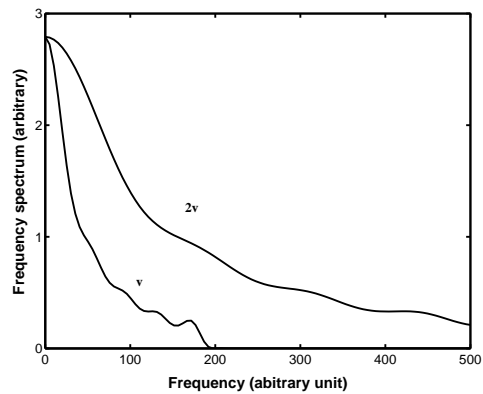}\smallskip
\caption{Frequency spectra for two rubbing speeds.} \label{fig2}
\end{figure}

To quantify our discussion, let us consider an example. Suppose
that $W=H$, $L=0.1W$, and $S$ is a constant. Two moving speeds are
considered: $v/c_s = 1/600$ and $1/300$. As an example, we compute
the sound field located at $(L,W/2,0)$. The frequency spectra for
the two moving speed are presented in Fig.~\ref{fig2}. The units
are arbitrary. Here it is clearly shown that there is indeed a
cut-off frequency at about 200 for the lower speed curve. Also for
the lower speed case, the discussed resonance feature does appear.
For the higher moving speed, the spectrum is broader. But the
resonance tends to be weaker. And the cut-off frequency is larger
(not shown). Compared to the lower speed case, the frequency band
is obviously broader. All these support the above qualitative
analysis.

In summary, a model is established to investigate the sound
generated by rubbing two objects. In the model, we have assumed
that the contacting surfaces are rough and the roughness is
described as completely random. The dependence of the frequency
spectrum on the moving speed and sample geometries is discussed.
The results seem to be in line with intuition or experience.

\end{document}